# SocRipple: A Two-Stage Framework for Cold-Start Video Recommendations

Amit Jaspal
Meta Platforms, Inc.
Menlo Park, CA, USA
ajaspal@meta.com

Kapil Dalwani
Meta Platforms, Inc.
Menlo Park, CA, USA
kapild@meta.com

Ajantha Ramineni
Meta Platforms, Inc.
Menlo Park, CA, USA
aramineni@meta.com

## ABSTRACT

Most industry-scale recommender systems face critical cold-start challenges—new items lack interaction history, making it difficult to distribute them in a personalized manner. Standard collaborative filtering models underperform due to sparse engagement signals, while content-only approaches lack user-specific relevance. We propose SocRipple, a novel two-stage retrieval framework tailored for cold-start item distribution in social graph-based platforms. Stage 1 leverages the creator's social connections for targeted initial exposure. Stage 2 builds on early engagement signals and stable user embeddings—learned from historical interactions—to "ripple" outwards via K-Nearest Neighbor (KNN) search. Large scale experiments on a major video platform show that SocRipple boosts cold-start item distribution by +36% while maintaining user engagement rate on cold-start items, effectively balancing new-item exposure with personalized recommendations.

## 1 Introduction

Modern industrial recommenders rely on embedding-plus-MLP architectures because they scale to billions of users and items while capturing complex interaction patterns [1, 2]. However, such models are data-hungry: when a brand-new item appears, the system cannot learn a reliable representation, leading to the classic item cold-start problem [3, 4]. In practice, the Pareto effect amplifies the issue: a tiny fraction of popular, well-trained items dominates impressions, reinforcing popularity bias and stifling catalog diversity [5].

Prior approaches like CLCRec tackle cold-start problem by learning contrastive representations from item content [6], while DropoutNet uses feature dropout to simulate cold-start during training [7]. While effective to a degree, these methods rely on static content signals and do not adapt to real-time user behavior or social influence [12, 13]. Graph-based methods propagate signals over user–item or social graphs via GNNs [8, 9] or friend-based regularization. Early-seed approaches push fresh items to a creator's followers [14, 15], but stop short of systematic, large-scale evaluation. None of these solutions sequentially fuse (i) creator-centric social seeding and (ii) real-time, embedding-driven neighbor expansion. Consequently, they either lack the high-precision bootstrap needed for the coldest items or the high-recall diffusion required to reach a broad yet relevant audience once early signals arrive.

To overcome these limitations, we propose SocRipple, a novel two-stage retrieval framework specifically designed for cold-start video recommendation in large-scale systems. Our approach is guided by the intuition that a creator's immediate social connections (e.g., friends, followers etc.) often provide the highest-precision initial audience for new content. SocRipple first leverages this social proximity for targeted seeding. Once initial engagement signals are gathered, it dynamically transitions to a second stage: embedding-based neighbor expansion. This stage utilizes stable, pre-computed user embeddings (learned via models like Two Tower [1]) and real-time K-Nearest Neighbor (KNN) search to efficiently propagate the new video to users with similar latent interests, guided by the early engagers. This cascading design allows fresh content to "ripple" outwards – from a high-confidence seed group to a wider, relevant audience – effectively balancing exploration (new item discovery) with exploitation (personalized recommendations). We demonstrate empirically on a large industrial dataset that SocRipple significantly boosts recall for new items while maintaining overall user engagement, showcasing a practical solution for managing the exploration-exploitation trade-off in cold-start scenarios.

## 2 Proposed Approach: SocRipple

SocRipple aims to efficiently distribute a new cold-start video $v_{new}$ uploaded by creator c by identifying relevant users $U$ over time. It leverages the social follower graph $G$ and pre-computed stable user embeddings $u_i \in R^d$ (learned via a base model trained on historical data.

### 2.1 Stage 1: Social Boost

Objective: Rapid, high-precision initial seeding and feedback collection.

Process:

1. **Identify and Ingest**: Upon creation of $v_{new}$ by creator $c$, quickly ingest for initial distribution.
2. **Retrieve Followers**: Get set of followers $F_c = u|(u,c) \in G$.
3. **Distribute**: Boost $v_{new}$ in recommendations for users in $F_c$.
4. **Logging**: Log initial positive engagement signals (e.g., views, likes) from $F_c$ in real-time. Let $U_{engaged_1} \subseteq F_c$ be the set of users who engaged positively.

### 2.2 Stage 2: Neighbor Expansion

Objective: To enrich target user's $u^t$ recommendations with relevant new videos by identifying content that similar users

(neighbors) have positively engaged with recently. This stage operates when a user requests their feed (Figure 1).

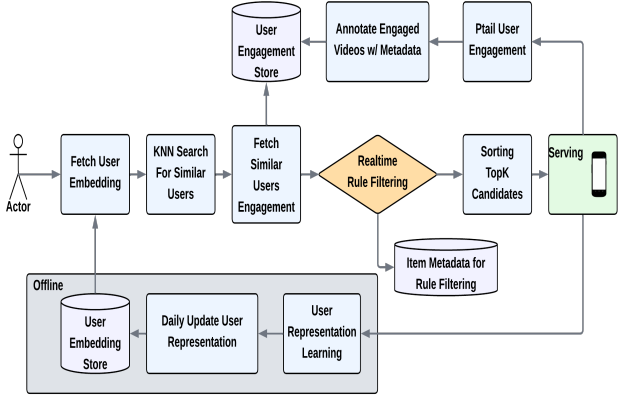

**Figure 1: End-to-end SocRipple system overview**

Process:
1. **Identify User's Neighbors**: Fetch the stable embedding $e(u^t)$ from the in-memory index and run a K-nearest-neighbor (KNN) search to obtain $N(u^t) = u_1 \ldots u_k$.
2. **Query Neighbor Engagement on New Videos**: For each neighbor $u_j$, look up the live engagement buffer. Collect the set $V_{new}(u_j)$ of cold-start videos that $u_j$ has positively engaged with within the last 24 hours.
3. **Aggregate candidate videos**: Form the candidate set by taking the union of all new-video lists returned for every neighbor. For every video in candidate set we compute: 1) neighbor-support count (# of neighbors who engaged), 2) similarity between $u^t$ and supporting neighbors, 3) freshness i.e. time since creation.
4. **Filter & selection**: Drop any video already shown to $u^t$. Apply lightweight rules and heuristics, finally rank remaining videos by a weighted score of (support count, neighbor similarity, freshness). In practice these weights are fine-tuned online via A/B test.
5. **Distribute**: Push the top ranked cold-start videos into ranking stage alongside candidates from other retrieval sources.

## 2.3 User Representation Learning

We employ a two-tower neural architecture to learn user embeddings from historical engagement logs. The model is trained using positive-only interactions (e.g., likes), each user $u$ and item $i$ is encoded into latent vectors $e_u, e_i \in R^d$, and their affinity is modeled by a dot product:

$$s(u,i) = e_u^T e_i \quad (1)$$

To optimize for engagement prediction, we use a sampled softmax loss with in-batch negatives. Given a mini-batch $B$ containing positive user–item pairs, we define the probability that user $u$ engages with item i as:

$$P(y_{u,i} = 1 \mid u, i) = \frac{exp(s(u,i))}{exp(s(u,i)) + \sum_{i' \in B-i} exp(s(u,i'))} \quad (2)$$

The training objective minimizes the negative log-likelihood over all positive instances:

$$\mathcal{L} = -\sum_{(u,i) \in B} \log P(y_{u,i} = 1 \mid u, i) \quad (3)$$

After training, the user tower is executed daily in an offline batch pipeline to compute stable user embeddings. These are stored in a FAISS approximate nearest neighbor (ANN) index, enabling fast retrieval of similar users in Stage 2 of SocRipple (Figure 1). This design allows new content to be efficiently propagated from early engagers to semantically aligned neighbors in real-time.

## 3 Offline Experiments

To evaluate the effectiveness of SocRipple framework, we conduct offline experiments on a large-scale dataset obtained from users' engagement on our video platform. We focus on users' engagement with cold-start videos, specifically items created in the last 6, 12, 24 hours. We measure retrieval performance by instrumenting standard recall@200 metric for cold start items [17]. Our experiments answer three research questions -- (RQ1) How much does SocRipple increase cold-start recall compared to traditional models? (RQ2) What is the incremental gain from neighbor expansion (Stage 2) using stable user representations learnt from engagement data? and (RQ3) How sensitive is SocRipple's performance to neighbor expansion hyperparameters?

### 3.1 (RQ1) Impact of SocRipple

To evaluate RQ1, we benchmarked SocRipple against three widely used cold-start retrieval baselines:

- DropoutNet [7], which combines collaborative filtering (CF) and content-based representations by feeding both into a multi-layer perceptron (MLP) and randomly dropping the CF component during training
- Content-KNN, a nearest-neighbor baseline that retrieves items based solely on pretrained content embeddings (e.g., metadata, tags) using cosine similarity.
- Item-KNN [10]. This is the standard item-based collaborative filtering method.

As shown in Table I, SocRipple significantly outperforms all baselines across all cold-start buckets. Notably, its relative gain is largest for the freshest items (≤ 6 hours), where conventional CF-based models struggle due to the absence of interaction history. The performance gap between SocRipple and CF baselines narrows as item age increases—consistent with the intuition that CF models become more effective as more engagement signals accumulate. This validates SocRipple's strength in handling true cold-start scenarios by avoiding dependence on aggregated item-level interactions and leveraging social graph of item distribution.

### Table 1: Cold start Recall@200 by item age

| Test Variant | Recall@200 Item age <= 6 hrs | Recall@200 Item age <= 12 hrs | Recall@200 Item age <= 24 hrs |
|---|---|---|---|
| DropoutNet | 5.8% | 7.2% | 8.8% |
| Content-KNN | 4.5% | 5.2% | 4.8% |
| Item-KNN | 3.2% | 5.4% | 7.5% |
| SocRipple | 12.8% | 13.1% | 13.2% |

## 3.2 (RQ2) Impact of user embeddings

To isolate the impact of Stage 2's embedding-based expansion, we compared full SocRipple against two variants:

- Stage 1 Only: New items recommended solely to the creator's direct social connections.
- Stage 1 + Social Graph Expansion (SGE): Stage 1 followed by expansion via the viewer's social graph, instead of embedding similarity.

This comparison disentangles the gains from embedding-based retrieval versus purely social graph propagation. As shown in Table 2 (for items ≤ 24 hours old), Stage 1 alone yielded 4.8% recall@200. Adding SGE modestly improved this to 6.7%. In contrast, full SocRipple, leveraging embedding-based expansion, achieved 13.2% recall@200—nearly double that of SGE.

These results confirm that embedding-based expansion accesses richer, orthogonal signals from shared behavioral preferences (interest-based similarity) rather than just declared connections (social ties). This enables SocRipple to dramatically widen discovery and accelerate new item reach beyond what social graph propagation alone can.

### Table 2: Neighbor Expansion Ablation (Recall@200)

| Neighbor Expansion Strategy | Recall@200 Item age <= 24 hrs |
|---|---|
| Stage 1 | 4.8% |
| Stage 1 with SGE | 6.7% |
| Stage 1 + Stage 2 | 13.2% |

## 3.3 (RQ3) Impact of neighbor expansion hyperparameters

To assess SocRipple's sensitivity to neighbor expansion in Stage 2, we varied two key hyperparameters: the number of neighbors (K), which controls expansion breadth of the user neighborhood in embedding space, and the number of cold-start videos per neighbor (M), which determines expansion depth per neighbor into candidate content space. Recall@200 improved steadily as K increased from 10 to 70, peaking at 0.13, but declined slightly beyond K=70 due to noise from distant neighbors. Similarly, increasing M up to 20 boosted recalls, after which performance plateaued and then regressed, indicating diminishing returns (Figure 2). We set M=20 based on these findings, which underscore the importance of tuning expansion parameters to balance recall and precision.

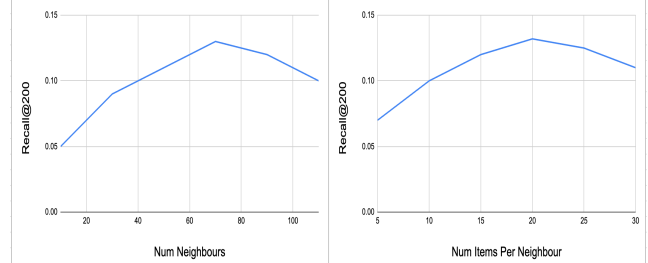

Figure 2: Recall@200 vs neighbor-expansion breadth and depth

## 4 Online Deployment

To evaluate the real-world performance of the SocRipple framework, we conducted a large-scale online A/B test over a two-week period on the video recommendation surface of our production platform. SocRipple was deployed as an additional candidate retrieval source alongside the existing collaborative filtering sources mentioned in the section above. Integrating this two-stage process, particularly the real-time neighbor expansion (Stage 2) involving on-the-fly KNN lookups and engagement aggregation, into our high-throughput production environment necessitated careful engineering and system optimizations to meet stringent latency requirements for candidate generation. The goal of this deployment was to assess SocRipple's impact on cold-start item discovery and overall user engagement. The reported improvements in engagement metrics are statistically significant.

We report improvements across three key metrics:

- **Topline engagement lift**: SocRipple led to a +0.22% increase in our primary engagement metric (proprietary), which correlates strongly with long-term user retention.
- **Improved cold-start coverage**: The distribution share of items less than 24 hours old in the treatment group increased by +36%, highlighting SocRipple's ability to improve early visibility of newly uploaded content. This resulted in thousands of additional new videos reaching meaningful impressions each day.
- **Stable engagement on cold-start items**: Despite increasing the exposure of new and previously unseen items, user engagement rates on cold-start content remained on par with the control group indicating that the additional items introduced were relevant and well-aligned with user interests.

These findings demonstrate that SocRipple effectively balances exploration of new content with preservation of engagement, making it well-suited for production deployment on dynamic video platforms.